\documentstyle[twoside,fleqn,espcrc2]{article}

\newcommand{\noi}{\noindent}
\newcommand{\eq}{\begin{equation}}
\newcommand{\en}{\end{equation}}
\newcommand{\eqa}{\begin{eqnarray}}
\newcommand{\ena}{\end{eqnarray}}

\newcommand{\lsim}{\raise0.3ex\hbox{$<$\kern-0.75em\raise-1.1ex\hbox{$\sim$}}}
\newcommand{\NT}{N_\tau}\def\nt{\ifmmode\NT\else$\NT$\fi}
\newcommand{\NS}{N_\sigma}\def\ns{\ifmmode\NS\else$\NS$\fi}

\newcommand{\AmS}{{\protect\the\textfont2
  A\kern-.1667em\lower.5ex\hbox{M}\kern-.125emS}}

\newcommand{\NP}{Nucl.\ Phys.}

\title{
\vspace{-1.8cm}
\hbox{}
{\small NOVEMBER 1993}  \hfill {\small HU BERLIN--IEP--93/5}    \break
\hbox{}                 \hfill {\small BIELEFELD BI-TP--93/63}   \break
                                                                  \break
The analysis of Polyakov loop and spin correlators in
              finite volumes
\thanks{TALK GIVEN AT THE LATTICE '93 INTERNATIONAL SYMPOSIUM
       LATTICE FIELD THEORY, DALLAS, USA, OCTOBER 12--16, 1993}
        \thanks{Work
       supported by the Deutsche Forschungsgemein-\break schaft}
       }

\author{J. Engels\address{Fakult\"{a}t f\"{u}r Physik, Universit\"{a}t
Bielefeld, 33615 Bielefeld, Germany}, %
  V. K.~Mitrjushkin\address{Fachbereich Physik, Humboldt-Universit\"{a}t,
10099 Berlin, Germany}%
\thanks{Permanent address:
Joint Institute for Nuclear Research, Dubna, Russia}
       and
T. Neuhaus$\mbox{}^{\mbox{\scriptsize a}}$
        }
\begin{document}

\begin{abstract}
 We derive an analytic expression for point to point correlation
functions of the Polyakov loop based on the transfer matrix formalism.
The contributions from the
eigenvalues of the transfer matrix including and beyond the mass gap
are investigated both for
the $2d$ Ising model and in finite temperature $SU(2)$ gauge theory.
We find that the
leading matrix element shows similar scaling properties in both models.
Just above the critical point we obtain for $SU(2)$ a Debye
screening mass $~\mu_D/T\approx4~$, independent of the volume.

\end{abstract}

\maketitle

 \section{Introduction}
%\section*{}
The determination of the correlation length $\xi$ and the screening
mass $\mu_D$ from point-to-point correlation functions of the
Polyakov loop is a non-trivial task, especially close to the critical
point of lattice gauge theories. The difficulties are resulting on
one hand from finite volume effects due to the nearby transition
and on the other hand from the unknown parametrisation of the heavy
quark potential in the non-perturbative regime.

{}~In the transfer matrix (TM) formalism the levels of the transfer matrix
provide an access to both $\xi$ and $\mu_D$ without the introduction
of an ansatz for the quark potential. The levels and matrix elements
may be obtained easily from fits to measured plane-plane ( or zero
momentum ) correlation functions since their TM form is known and
simply exponential. In this paper we intend to derive
the corresponding  expression for point-to-point correlation functions.
 In the $2d$ Ising
 model we test the validity of our TM formula by comparison to the
 results obtained from plane-plane correlators.
 Simultaneously we are
able to determine, where levels beyond the mass gap are of importance
and what can be expected from such an analysis.

 \section{Correlation functions in the transfer matrix formalism}
 Let us consider $d-$dimensional spatial lattices of size $N^{d-1}L$,
where $N$ denotes the number of points in each transverse direction
and $L$ that in one selected direction ( the $z-$direction ).
The lattice spacing $a$ is set to unity in the following. The partition
funtion is then

\eq
Z \equiv \mbox{Tr} \left( {\bf V}^{L} \right),
\en

\noi and $~{\bf V}~$ is the transfer matrix in $~z-$ direction.
Its eigenstates $~\mid \! n \rangle~$  and
eigenvalues $~\lambda_{n}~(n=0,~1,~2,~...)~$

\eq
{\bf V} \mid \! n \rangle = \lambda_{n} \cdot \mid \! n \rangle~;
\quad \lambda_{n} \equiv e^{-\mu_{n}} ;
                                                    \label{lambd}
\en

\eq
\mu_{0} < \mu_{1} < \mu_{2} < ...~.
\en
are chosen to be orthonormal. In addition we normalize our partition
function such that we have for the vacuum state

\eq
\lambda_0 = 1,~\mu_0 = 0 .
\en
Next we define zero momentum operators by

\eq
 \tilde {{\cal O}}(z) = N^{-\frac{d-1}{2}} \! \cdot
 \sum_{\vec{x}_{\perp}} {\cal O}(\vec{x}_{\perp},z) ~,
                                                \label{otilde_is}
\en
where
\noi $~{\cal O}(\vec{x}_{\perp},z)~$ is the Polyakov loop
$~{\cal P}(\vec{x}_{\perp},z)~$ for the $3+1$ dimensional $~SU(2)~$
gauge theory and the spin $~\sigma_{x,z}~$ for the $2d$ Ising model.
The corresponding correlation functions are

\eq
\tilde{\Gamma} (z)  =
\langle \tilde {{\cal O}}(z) \cdot \tilde { {\cal O}}(0) \rangle,
\en
resulting in
\eq
\tilde{\Gamma} (z)  =
\sum_{n < m} {c_{mn}^2 \over Z} e^{-\mu_n L}
\Bigl[ e^{- \mu_{mn} z}
+ e^{- \mu_{mn} (L-z)} \Bigr],
                                                 \label{cor1_is}
\en
where
\eq
\mu_{mn} = \mu_m - \mu_n ;
{}~c_{mn} =
\langle n \! \mid \tilde{{\cal O}}(0) \mid \! m \rangle~,
\en
are the level difference and the transition matrix element.
Due to the symmetry properties of the eigenstates under
tranformations, which change the sign of ${\cal O}$, $~c_{nn} = 0~$.

Below the critical point $~\beta < \beta_{c}~$ the lowest nonzero
energy level $~\mu_{1}~$ ( the mass gap )
defines the large distance behaviour of the correlator.
We therefore define the correlation length at
$~\beta \stackrel{<}{\sim} \beta_{c}~$ by
\eq
\xi_{-}(\beta) \equiv \mu_{1}^{-1}
\sim \mid \! \beta - \beta_{c} \! \mid ^{-\nu}.
\en

At $\beta > \beta_c$ the mass gap $\mu_1 \approx 0$ and the large
distance behaviour is given by the next level difference
$\Delta\mu$, so that the Debye mass is
\eq
\mu_D = 2m_D \equiv \Delta\mu~,
\en
where $m_D$ is the perturbative screening mass.

A similar formula as eq.\ref{cor1_is} may now be found for the
point-to-point correlator
\eq
\Gamma (\vec{x})  =
\langle {\cal O}(\vec{x}) \cdot  {\cal O}(0) \rangle,
\en
in the following way. The Fourier transforms of $\tilde{\Gamma} (z)$
and $\Gamma (\vec{x})$ are related by
\eq
\tilde{\Gamma}(p_{z}) \equiv \Gamma(\vec{p}_{\perp}=0,p_{z})~.
                                                         \label{cor3_is}
\en
This leads us to the ansatz

\eq
\Gamma (\vec{p}) = Z^{-1} \sum_{n < m} c_{mn}^{2}e^{-\mu_{n} L}
 G(\vec{p}; \mu_{mn})~,
                               \label{Gamma}
\en

\noi where

\eq
G(\vec{p}; \mu) = {
2 \left(1-e^{-\mu L}\right) \sinh \mu \over
4 \sinh^{2} \frac{\mu}{2} +
                  \sum_{i=1}^{d} 4 \sin^{2} \frac{p_{i}}{2}~
       }~,
                                           \label{ansatz}
\en
and $G(\vec{p}_{\perp}=0,p_z; \mu )$ is just the Fourier transform of
$\Bigl[ e^{- \mu_{mn} z} + e^{- \mu_{mn} (L-z)} \Bigr]$, i.e. we
have added corresponding contributions for the missing momenta
components in the denominator of eq.\ref{ansatz}.
It is now straightforward to obtain $\Gamma(\vec{x})$ by
another Fourier transformation of eq.\ref{Gamma}.

\section{Results}

We have used the $2d$ Ising model to test our ansatz, eq.\ref{Gamma}.
To this end we have measured plane-plane and point-to-point correlators
on $N=L=30,40,50,60$ lattices. At each point $500000$ cluster updates
were performed and measurements taken every 10th update. In the
twodimensional model the levels $\mu_n$ are explicitly known
\cite{o,sml}. We have carried out fits to both correlators with
varying numbers of levels to obtain the matrix elements. Both formulae
lead to the same results, when the maximal number of levels is taken
into account which lead to non-negative $c^2_{mn}$; i.e. our ansatz
is definitely confirmed. The final result is shown in Fig.1 for
$N=L=30$. We find that for $\beta < \beta_c$ only one term with
 $\mu_{10}=\mu_1$, the mass gap, contributes; near $\beta_c$ up to
three terms are essential and well above $\beta_c$, where $\mu_1 \approx
0$ only one more term is present.
\begin{figure}[htb]
%\framebox[63mm]{\rule[-11mm]{0mm}{52mm}}
\vskip 57mm
\caption{
The lowest level differences $~\mu_{1},~\mu_{3}$ and $~
\mu_{21}$ and the respective matrix elements
$~c_{10}^2,c_{30}^2~$ and $~c_{21}^2e^{-\mu_{1}L}~$
vs. $~\beta/\beta_c~$ in the $~2d$ Ising model.}
\label{fig:Ising}
\end{figure}

\begin{figure}[htb]
%\framebox[63mm]{\rule[-11mm]{0mm}{52mm}}
\vskip 57mm
\caption{
$N\mu_1$ resulting from one- and two-level fits (open
and filled symbols) on $N=18,26$ (squares and diamonds) lattices
for $SU(2)$. The inset shows for $N=26$ also 3-level fits (circles).
The dotted line indicates $\beta_c$.
}
\label{fig:fitmu1}
\end{figure}

\vspace{-0.3cm}
We now want to apply our TM formula for the point-to-point correlator
to $SU(2)$ gauge theory. The Monte Carlo data \cite{Oldpa}
 were computed on $\NS^3 \times 4$
lattices, $\NS = 12,18,26$  with $10^5-4\cdot10^5$
updates and measurements every 10th sweep.

 Here the level differences are unknown and have to be determined
through the fit. In general we find a very similar behaviour as in
the $2d$ Ising model. Fits with more than two levels are only possible
on the largest lattice very close to the transition. Otherwise one
either obtains negative squares of matrix elements or there is no
minimum of $\chi^2$. Taking into account more terms in eq.\ref{Gamma}
for the fits tends to decrease the result for the mass gap level.
This is shown in Fig.2.

It is interesting to look at the behaviour of the next to leading
level ( or level difference ) $\mu_2$.
%, which we eventually want to
%identify with the Debye screening mass $\mu_D$ above the critical
%point.
 As can be seen from Fig.3, $\mu_2$ drops from a higher value
below $\beta_c$ at the transition to a value near to one ( in lattice
units ) and stays then relatively constant and moreover independent
of the lattice sizes used here. This second level fixes the large
distance behaviour above $\beta_c$
of the correlation functions, since $\mu_1$, as
is evident from Fig.2, is essentially zero there and a third level
does not contribute. Therefore we identify it with $\mu_D$.
Because we have $\NT=4$ we are led to
a ratio $\mu_D/T\approx 4$, slightly higher than found with conventional
methods\cite{Eng2}.

It can be shown \cite {Newpa}, that for $N \rightarrow \infty$
\eq
     ~c_{mn}^{2} \sim N^{0}~; ~\rm{for}~ \beta < \beta_c,
\en
\eq
     ~c_{10}^{2} \sim N^{d-1};~\rm{for}~ \beta > \beta_c,
\en
and from finite size scaling theory \cite{Oldpa}
for $\beta \approx \beta_c $
\eq
     ~c_{10}^{2} = N^{\gamma/\nu-1} f(xN^{1/\nu}).
                                       \label{scaling}
\en
These scaling properties are all well confirmed by both the
$2d$ Ising model and the $SU(2)$ gauge theory results.

\begin{figure}[htb]
%\framebox[63mm]{\rule[-11mm]{0mm}{52mm}}
\vskip 57mm
\caption{
The second level $~\mu_2~$ from two-level fits
in $~SU(2)~$ gauge theory for $~N=12,18,26~$~(crosses,squares,diamonds)
vs. $~\beta~$.
}
\label{fig:fitmu2}
\end{figure}

\end{document}